\documentclass[%
prb,
10pt,
amsmath,
amssymb,
]{revtex4}

\usepackage{titlesec}
\titlespacing*{\section}
{0pt}{1.5ex}{1.5ex}
\titlespacing*{\subsection}
{0pt}{1ex}{1ex}
\titlespacing*{\subsubsection}
{0pt}{1ex}{1ex}
\usepackage{setspace}
\usepackage{tabularx}

\usepackage[version=3]{mhchem} 
\usepackage[dvipsnames]{xcolor}
\usepackage{amsfonts}
\usepackage{graphicx}
\usepackage[colorlinks=true,citecolor=red]{hyperref}

\usepackage[dvipsnames]{xcolor}
\usepackage{url}
\usepackage{color}
\usepackage{dcolumn}
\usepackage{bm}
\usepackage[mathlines]{lineno}
\usepackage{comment}
\usepackage[version=3]{mhchem}
\usepackage[capitalise]{cleveref} 

\usepackage{graphicx} 
\usepackage{amsmath}

\usepackage{wrapfig,floatflt}

\renewcommand{\bm}[1]{\mathbf{#1}}
\newcommand*{\myfont}{\fontfamily{phv}\selectfont}
\DeclareTextFontCommand{\textmyfont}{\myfont}

\usepackage{concmath}

\let\oldbibliography\thebibliography
\renewcommand{\thebibliography}[1]{\oldbibliography{#1}
\setlength{\itemsep}{0pt}} 
\allowdisplaybreaks

\makeatletter
\renewcommand{\thesection}{\arabic{section}}
\renewcommand{\thesubsection}{\arabic{section}.\arabic{subsection}}
\renewcommand{\p@subsection}{}

\renewcommand{\p@subsubsection}{}
\makeatother
\usepackage{capt-of}
\usepackage{tabularx}

\begin{document}

\title{MACE Foundation Models for Lattice Dynamics: A Benchmark Study on Double Halide Perovskites}

\author{Jack Yang}
\email{jinaliang.yang1@unsw.edu.au}
\affiliation{School of Materials Science and Engineering, University of New South Wales, Sydney, New South Wales 2052, Australia}

\author{Ziqi Yin}
\affiliation{School of Materials Science and Engineering, University of New South Wales, Sydney, New South Wales 2052, Australia}

\author{Lei Ao}
\affiliation{Jiangxi Provincial Key Laboratory of Advanced Electronic Materials and Devices, Jiangxi Science and Technology Normal University, Nanchang 330018, China}
\affiliation{School of Physics, University of Electronic and Technology of China, Chengdu 610054, China}
\affiliation{School of Materials Science and Engineering, University of New South Wales, Sydney, New South Wales 2052, Australia}

\author{Sean Li}
\affiliation{School of Materials Science and Engineering, University of New South Wales, Sydney, New South Wales 2052, Australia}

\date{\today}

\begin{abstract}
Recent developments in materials informatics and artificial intelligence has led to the emergence of foundational energy models for material chemistry, as represented by the suite of MACE-based foundation models, bringing a significant breakthrough in universal potentials for inorganic solids. As to all method developments in computational materials science,  performance benchmarking against existing high-level data with focusing on specific applications, is critically needed to understand the limitations in the models, thus facilitating the ongoing improvements in the model development process, and occasionally, leading to significant conceptual leaps in materials theory. Here, using our own published DFT (Density Functional Theory) database of room-temperature dynamic stability and vibrational anharmonicity for $\sim2000$ cubic halide double perovskites, we benchmarked the performances of four different variants of the MACE foundation models for screening the dynamic stabilities of inorganic solids. Our analysis shows that, as anticipated, the model accuracy improves with more training data. The dynamic stabilities of weakly anharmonic materials (as predicted by DFT) are more accurately reproduced by the foundation model, than those highly anharmonic and dynamically unstable ones. The predominant source of error in predicting the dynamic stability arises predominantly from the amplification of errors in atomic forces when predicting the harmonic phonon properties through the computation of the Hessian matrix, less so is the contribution from possible differences in the range of the configurational spaces that are sampled by DFT and the foundation model in molecular dynamics. We hope that our present findings will stimulate future works towards more physics-inspired approaches in assessing the accuracy of foundation models for atomistic modelling.
\end{abstract}	

\maketitle
\onehalfspacing

\section{Introduction}

Atomistic modellings play a pivotal role in modern materials physics and chemistry, which is complementary to the experimental endeavours in discovering new materials for structural, electronic, energy harvesting and many other applications. Primarily, the key information to be extracted from atomistic modellings, particularly  DFT\cite{kohn1965self} (Density Functional Theory), which is the workhorse for modern computational materials science, is the total energy of a material with a specific atomistic structure. This information is of profound importance in materials discoveries, because it is one of the key indicators for materials' stabilities (and to some extent, synthesisabilities\cite{bartel2018physical,mcdermott2021graph}). From here, other physical properties, such as electronic structures, magnetic ground states and optical responses, can be acquired as auxiliaries to a DFT calculation, since it solves a good approximation to the fundamental physical equation that governs the quantum mechanical behaviours of electrons in materials.

One of the key drawbacks of DFT is its $\mathcal{O}(N^3)$ scaling behaviour to the system size measured as the number of atoms $N$, which makes it computationally very expensive to be applied in large-scale modellings, such as for chemically disordered high-entropy materials\cite{oses2020high}, and systems in which many-body interactions significantly dictate their physical properties\cite{giustino2017electron}. Traditionally, this bottleneck was overcome by employing atom-atom force fields\cite{harrison2018review} that are designed with fast-to-calculate analytical functions based on known physics (\emph{e.g} harmonic potential for bond stretching) with parameters fitted to a single or a specific set of material(s). However, this also imposes significant limitations in applying these tailored-made force fields to correctly model exotic materials behaviours such as anharmonic phonon vibrations and Coulomb interactions between polarised charge densities, thus restricting the model transferability to systems that have not been parameterised. This makes the development of universal force field become a long-time challenge in materials modelling. This, nevertheless, is not a problem for DFT. 

Over the past two decades, machine-learning interatomic potentials (MLIPs) have emerged and rapidly developed to bridge the gap between DFT- and force-field-based energy models. The early successes in this endeavour is largely rooted in using kernel regressions based on hand-crafted and physically inspired descriptors for local atomic environments\cite{bartok2010gaussian,jinnouchi2019fly}. This knowledge has been fueled into the recent development of deep-learning potential models such as Schnet\cite{schutt2018schnet}, NequIP\cite{batzner20223}, MACE\cite{batatia2022mace} and So3krates\cite{frank2022so3krates}. Some of them\cite{frank2022so3krates} have further incorporated deep-learning architectures, such as the attention mechanism\cite{vaswani2017attention} from the large language models to capture long-ranged atom-atom interactions in materials, demonstrating the cross-paradigm nature in this field of research. In the meantime, the continuous expansions of large computational materials databases, such as the Materials Project\cite{jain2013commentary} and OMAT\cite{barroso2024open} have provided the community with rich resources of materials structural, energetic and property data that are generated in a consistent level of theory. The scale and diversity of hundreds of millions of first-principles calculations provided by these databases unlock  our capabilities to develop a transferable universal foundation energy model for (solid-state) materials across a significant portion of the existing chemical space\cite{merchant2023scaling}. 

This significant milestone can be exemplified by the recent achievement behind the releases of a suite of foundation models\cite{batatia2023foundation} based on the MACE\cite{batatia2022mace} (Message-passing Atomic Cluster Expansion) architecture, which is the focus of this study. More specifically, to learn the total atomic interaction energies in chemical systems, MACE combines the graph neural network\cite{duval2023hitchhiker} that models chemical structures as graphs and utilises the message-passing mechanism\cite{gilmer2017neural} to exchange chemical bonding information across multiple message-passing layers in the network, together with the atomic cluster expansion\cite{drautz2019atomic} formalism to ensure the equivariance of the local atomic environment is preserved as the messages are passed through the network. 

In the first release\cite{batatia2023foundation}, dubbed as \texttt{mp-0}, the foundation model was trained on the \texttt{MPtrj}\cite{deng2023chgnet} dataset, which contains a large number of static calculations and structural optimisation trajectories  for inorganic solids at the PBE$+U$  level of theory. This includes approximately 1.5M structures with 90\% of them of less than 70 atoms per unit cell. With this level of coverage, the \texttt{mp-0} model had been applied to demonstrate its applicability to 30 different categories of atomistic simulations, ranging from ice structures, metal organic frameworks, heterogeneous catalysts, amorphous structures, to complex liquid-solid interfaces. 

However, even at this training scale, the intrinsic problem associated with any MLIPs cannot be overlooked in the foundation model, that is, at its best, the model accuracy can only be as good as the underlying theory that was applied to generate the training data. This issue has already been addressed\cite{batatia2023foundation}, for example, the DFT setting for generating the \texttt{MPtrj} dataset is less tight compared to that required for accurate phonon calculations, as such, the error in reproducing the DFT phonon bandwidth with \texttt{mp-0} is $\sim$1-2 THz, that is an order of magnitude larger than the results from highly specialised model\cite{george2020combining}. Overcoming such a shortage is undoubtedly a key driving force for the ongoing improvement of the MACE-foundation models (\cref{tab:mace-models}). This is because many key physical properties of materials, such as dynamic stabilities\cite{stoffel2010ab,monacelli2021stochastic}, electron dynamics and superconductivities\cite{giustino2017electron}, all share strong link to the phononic behaviours of the materials. A notable improvement in predicting phonon properties is expected with the latest iteration of the model that was trained on the \texttt{OMAT}\cite{barroso2024open} database. 

\begin{table}[tbh]
    \centering
    \caption{Overviews of the MACE foundation models that are benchmarked in this work.}
    \label{tab:mace-models}
    \begin{tabular}{p{4cm}p{2cm}p{3cm}p{3cm}p{3cm}}
    \hline\hline
       \textbf{Model Name}  & \textbf{Elements Covered} & \textbf{Training Dataset}  & \textbf{Level of Theory}  & \textbf{Notes}\\
       \hline
       \texttt{matpbs-pbe-omat-ft}  & 89  & \texttt{MATPES-PBS}\cite{kaplan2025foundational}   &  DFT (PBE) & No $+$U correction\\

       \texttt{mpa-0-medium} & 89 & \texttt{MPtrj}\cite{deng2023chgnet}+\texttt{sAlex}\cite{mehl2017aflow} & DFT (PBE+U) & Improved accuracy particularly high pressure stabilities \\

       \texttt{mp-0b3-medium} & 89 & \texttt{MPtrj} & DFT (PBE+U) & Improved phonon properties\\

       \texttt{omat-0-medium} & 89 & \texttt{OMAT}\cite{barroso2024open} & DFT (PBE+U) & Excellent phonon properties\\
    \hline\hline
    \end{tabular}
    
\end{table}

This is an interesting development, as the major improvement from \texttt{MPtrj} to \texttt{OMAT} was not necessarily on DFT setting that improved the phonon  accuracy, but an expansion in the dataset size which contains, for example, rattled crystal structures sampled from Boltzmann distributions as well as molecular dynamic trajectories.  This highlights the importance of including more training data that can closely trace the topologies of the underlying DFT potential energy surfaces (PES) for different materials as a key strategy for developing foundational models for materials chemistry. 

A particularly relevant case is anharmonic\cite{knoop2020anharmonicity} solids, for which vibrating atoms tend to traverse a PES with complex topology that notably deviates from the idealised parabolic shape. A representative material system is the cubic perovskites, for which the high-symmetry cubic structure is a saddle point on a double-well-shaped PES, that can be expressed as a fourth-order polynomial\cite{yang2020composition}. Solving the eigenvalue equation for the dynamical matrix of harmonic phonons for these systems typically leads to imaginary phonon frequencies\cite{pallikara2022physical} at the high symmetry points in the reciprocal space, which correspond to the structurally-related antiferrodistortive\cite{klarbring2018nature} or electronically-related ferroelectric\cite{bersuker2013pseudo} instabilities. Distorting the high-symmetry cubic perovskite structure along the eigenvectors of these imaginary phonon eigenvectors corresponds to symmetry-breaking events that will drive the structure into an energetically more stable state. Physically, the depth of the double-well potential plays a strong contribution towards the degree of vibrational anharmonicities. The latter is strongly related to the chemical constituents and bonding characteristics of the materials.

The above idea inspires our present investigation, in which we use our unique harmonic phonon and room-temperature \emph{ab initio} molecular dynamics (AIMD) database of $\sim2000$ halide double perovskites (HDPs)\cite{yang2022mapping}, covering a diverse range of materials' dynamic stabilities and vibrational anharmonicities\cite{knoop2020anharmonicity} while maintaining structural homogeneity (all being with the $Fm\bar{3}m$ space group symmetry), to benchmark the performances of the MACE foundation models (\cref{tab:mace-models}) in tracing the topologies of PES across a range of different degrees of anharmonicity.

Our detailed analysis reveals the followings. The previously established anharmonicity score\cite{knoop2020anharmonicity} is fundamentally equivalent to a measurement of force-fitting residue\cite{yang2022atlas}, which can be used to reveal (a) part of the chemical space where the foundational models performed well (poorly) in reproducing the DFT-PES, as well as (b) regions of the DFT-PES for an individual material that are well (poorly) reproduced by the foundation model. Overall, it shows that highly anharmonic part of the chemical space and the DFT-PES for individual material are generally less well reproduced by the foundation model. Nevertheless, if both the harmonic and anharmonic contributions to the atomic forces are computed consistently with the same  energy model, it should provide a reasonably good indication to the dynamic stabilities of a material that is quantitatively aligned with the DFT result, suggesting these foundation models\cite{batatia2023foundation} are indeed sufficient for accelerating large-scale screening of finite-temperature materials stabilities, which is a critical component in the theory-driven materials discoveries.

\section{Methodologies}
\label{sect:method}
\noindent
\textbf{HDP database} For the details of DFT calculations that are used to generate the database, as well as the chemical space covered, we refer the readers to our original publication\cite{yang2022mapping}. Details for accessing the database are provided in \cref{sect:datalink}. All DFT calculations were performed at the PBE (Perdew-Burke-Ernzerhof)\cite{perdew1996generalized} level of theory without Hubbard $U$ correction, which is broadly consistent with the parameterisation level of the MACE foundation models (\cref{tab:mace-models}). With respect to the current work, the most important DFT data for each HDP that is contained in this database includes:
\begin{enumerate}
    \item Harmonic force constant matrix $\mathbf{\Phi}_{ij}^{\alpha\beta}$ computed from the finite-displacement approach in real space\cite{togo2015first}. Physically, each matrix element of the force constant matrix corresponds to the force appears to be on atom $i$ along the $\alpha$ Cartesian direction when atom $j$ is displaced along the $\beta$ direction. The availability of the force constant matrix enables us to surrogate a $(3N+1)$-dimensional (with $N$ being the number of atoms in the simulation supercell) harmonic approximation to the PES in the vicinity of the local minimum that corresponds to the high-symmetry $Fm\bar{3}m$ structure of HDP. Diagonalising the Fourier transformation of $\mathbf{\Phi}$ will provide us with the phonon eigenfrequencies $\{\omega(\bm{q},n)\}$, where $\bm{q}$ is the phonon wavevector in the first Brillouin zone, and $n$ is the band index for a given $\bm{q}$. The phonon dispersion relationship can be obtained by connecting $\{\omega(\bm{q},n)\}$ with the same $n$ across all symmetrically unique $q$-points in the first Brillouin zone, from which one can compute the corresponding phonon group velocities as $\bm{v}_{g}(\bm{q},n)=\partial \omega(\bm{q},n)/\partial \bm{q}$.
    \item AIMD trajectory which contains a set of time-dependent atomic coordinates and forces $\{\bm{R}(t),\bm{F}(t)\}$ that are sampled at 300 K for up to 1.6 ps at 1 fs time step under the $NVT$ ensemble. AIMD simulations enable us to sample a wider (higher-energy) portion of the energy basin that is centred around the $Fm\bar{3}m$ local minimum. Since DFT does not take any assumption on the topology of the underlying PES (as opposed to traditional force fields), but solely determines the local PES gradient (encapsulated in $\bm{F}(t)$) by solving the electronic structure at the given atomic configuration $\bm{R}(t)$, it is able to capture the nonparabolic aspect in the topology of the PES, especially at distant to the local minimum. 
\end{enumerate}

\noindent
\textbf{Anharmonicity score} By combining the information of harmonic force constants and AIMD trajectories, \citet{knoop2020anharmonicity} proposed the following score to measure the degree of vibrational anharmonicity of a material at a given temperature $T$:
\begin{equation}\label{eq:anharmonic_score}
    \sigma^{(2)}=\sqrt{\sum_{i,\alpha}\left\langle \left(F^{\alpha,A}_{i}\right)^2 \right\rangle_{T}\Big/ \sum_{i,\alpha}\left\langle \left(F_{i}^{\alpha}\right)^2 \right\rangle_{T}}.
\end{equation}
Essentially, the anharmonicity is measured by comparing the standard deviation of the total ($F$) and anharmonic ($F^{A}$) atomic forces sampled across the AIMD trajectory. Here, $F_i^\alpha$ ($F_i^{\alpha,A}$) is the total and anharmonic force on the $i$-th atom in the simulation cell along the $\alpha$-Cartesian direction, and they are related to each other via $F_i^{\alpha,A}=F_i^\alpha-F_i^{\alpha,(2)}$, in which the harmonic component of the atomic force can be computed from the force constant $\mathbf{\Phi}_{ij}^{\alpha\beta}$ as $F_i^{\alpha,(2)}=-\sum\limits_{{j,\beta}}\mathbf{\Phi}_{ij}^{\alpha\beta}\bm{u}_i^\alpha$, with $\bm{u}_i^\alpha$ being the atomic displacement from its equilibrium position. Summation over all atoms and three Cartesian directions for a given AIMD frame gives the time-dependent $\sigma^{(2)}(t)$, which provides a measure of anharmonicity for the particular  frame, whereas taking the average $\langle\sigma^{(2)}(t)\rangle_t$ over the entire AIMD trajectory provides a single numerical measure of the anharmonicity of a given material at $T$. The later also determines the mechanical stability of the materials, as those with $\langle\sigma^{(2)}(t)\rangle_t>1$ are considered as unstable at the simulated temperature $T$.  

There are two important aspects of the anharmonicity score, which is rooted from its definition \cref{eq:anharmonic_score}. Firstly, as the harmonic force is directly proportional to the atomic displacements, $\sigma^{(2)}(t)$ can be treated as a single-valued proxy to gauge the range of the phase space being sampled in an MD simulation\cite{yang2020mapping}. Secondly, \cref{eq:anharmonic_score} shows that the anharmonicity score is fundamentally a measure of standard deviation, which is also a measure of force fitting accuracy in all MLIP developments\cite{yang2022atlas}, hence, as shall be shown below, it can provide us with more physical and diagnostic insight into the chemical and structural phase spaces in which the foundation models performed well or extrapolate poorly in practical simulations. 

\noindent
\textbf{Configurational Space Analysis} Unsupervised machine learning provides a powerful way to compare the configurational spaces sampled with two different energy models, in this case, DFT and MACE foundation model (more specifically, the \texttt{omat-0-medium} model), which will enable us to understand more deeply the discrepancies in the dynamic stabilities of HDPs that are acquired from these two different energy models. For this purpose, we first mathematically encode each MD frame with the SOAP\cite{bartok2013representing} (Smoothed Overlap of Atomic Positions) structure descriptor. Technically, all atoms in the simulation supercells were included as the `centres' on which their surrounding atomic environments are considered in constructing the structure descriptor for the crystal, with the periodic boundary conditions taking into account. The radial cut-off distance for finding the neighbouring atoms to each centre is $r_{c}=5$ \AA. Each atom is modelled as a normal distribution centred at its Cartesian coordinates, with a standard deviation of $\sigma_{c}=0.1$ \AA.  The number of basis functions used to expand the radial and angular distribution of the atomic environment around each centre are set to $n_{\max}=7$ and $\ell_{\max}=6$, respectively. The REMatch\cite{de2016comparing} (Regularized Entropy Match) similarity metric is employed to compute the similarity between two multiatomic MD frames, from here, the similarity kernel, which encodes the pairwise structural similarities among all MD frames in the trajectory can be constructed. The SOAP-REMatch kernel is then subsequently used to construct a two-dimensional map with the Kernel Principal Component Analysis (KPCA), enabling us to visually compare the configurational spaces being sampled by AIMD and MACE-MD. The SOAP-REMatch kernel is computed using the \texttt{dscribe}\cite{himanen2020dscribe} package, and the KPCA analysis is performed with \texttt{scikit-learn}\cite{pedregosa2011scikit}.

\section{Results and Discussions}

\subsection{Harmonic Phonons}\label{sect:harmonic_phonons}

We first examine the performances of the MACE foundation models in reproducing the key phononic characteristics of solids computed from periodic DFT.  For this purpose, we first recompute the harmonic force constant matrix using the same finite-displacement approach with the same size of $(2\times2\times2)$ supercell for each HDP as in our previous work\cite{yang2022mapping}, except now the atomic forces on each finite-displaced supercell structure are computed with the MACE foundation models, from which the harmonic constant matrix $\Phi_{\mbox{\scriptsize{MACE}}}$ can be determined. As detailed in the Methodologies section, diagonalising $\Phi_{\mbox{\scriptsize{MACE}}}$ will give us a set of phonon eigenfrequencies $\{\omega_{\mbox{\scriptsize{MACE}}}(\bm{q},n)\}$ and group velocities $\{\bm{v}_{\mbox{\scriptsize{MACE}}}(\bm{q},n)\}$, from which the following two root-mean-squared-errors (RMSE) metrics were applied to gauge the deviation of the MACE predicted phononic properties from those computed with DFT: (1) RMSE in $\omega^2$, defined as 
\begin{equation}
    \label{eq:rmse_omega}
    \mbox{RMSE}(\omega^2)=\sqrt{\frac{1}{N_{\bm{q}}N_{n}}\sum_{\bm{q},n}\biggl\|\omega^2_{\mbox{\scriptsize{MACE}}}(\bm{q},n) -  \omega^2_{\mbox{\scriptsize{DFT}}}(\bm{q},n) \biggl\|},
\end{equation}
which eliminates the possible complication of comparing a real and an imaginary phonon eigenfrequency with the same combination of $\{\bm{q},n\}$, Here $N_{\bm{q}}$ is the total number of wavevectors sampled in the first Brillouin zone and $N_n$ is the total number of eigenstates for a given eigenvector $\bm{q}$. Physically, the magnitudes of the phonon eigenfrequencies provide good indications on the mechanical strengths of a solid. (2) RMSE in $\bm{v}_g$, defined as 
\begin{equation}
    \label{eq:rmse_vg}
    \mbox{RMSE}(\bm{v}_g)=\sqrt{\frac{1}{N_{\bm{q}}N_{n}}\sum_{\bm{q},n}\bigg(\bm{v}_{\mbox{\scriptsize{MACE}}}(\bm{q},n)-\bm{v}_{\mbox{\scriptsize{DFT}}}(\bm{q},n)\bigg)^2, } 
\end{equation}
which provides a good indication on reproducing the shape of the DFT-phonon dispersion relationship. 

\begin{figure}[tbh]
    \centering
    \includegraphics[width=\linewidth]{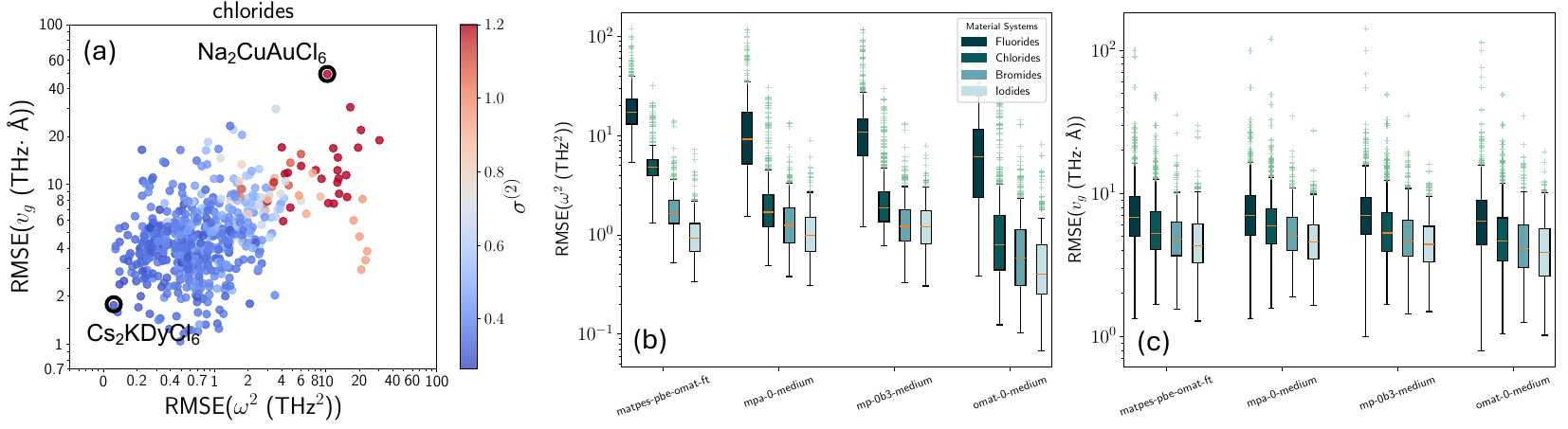}
    \caption{Accuracies of the MACE foundation models in predicting the harmonic phonon properties for HDPs. (a) Correlations between the RMSEs in predicting the phonon eigenfrequencies and group velocities with respect to the DFT results for chloride HDPs using the  \texttt{omat-0-medium} foundation model. Each data point is colour-coded according to its anharmonicity score $\sigma^{(2)}$ computed from DFT\cite{yang2022mapping}. Two extrema with the best (lower left) and worst (upper right) prediction accuracies are highlighted, the corresponding phonon dispersion relationships for which are shown in \cref{fig:phonon_dispersion_compare_omat}. (b) and (c) show the box plots that present the ranges of RMSEs in predicting the phonon eigenfrequencies and group velocities using all four foundation models for each groups of HDPs as categorised by the halide anion. The orange line indicates the medium RMSE. The box limits represent the 1st and 3rd quartiles. The whiskers show the range of the RMSEs within 1.5$\times$ the interquartile range of the box limits, and the outliers are denoted by light cyan plus symbols. }
    \label{fig:harmonic_phonon_summary_plots}
\end{figure}

As an example, \cref{fig:harmonic_phonon_summary_plots}(a) shows the relationship between $\mbox{RMSE}(\omega^2)$ and $\mbox{RMSE}(\bm{v}_g)$ for chloride HDPs computed with the \texttt{omat-0-medium} model. To showcase what these two error metrics reflect, we also show in \cref{fig:phonon_dispersion_compare_omat} the corresponding comparisons of the phonon dispersion curves computed with the \texttt{omat-0-medium} model and DFT for the two extrema (\ce{Cs2KDyCl6} and \ce{Na2CuAuCl6}) identified in \cref{fig:harmonic_phonon_summary_plots}(a). For \ce{Cs2KDyCl6} which has the lowest $\mbox{RMSE}(\omega^2)$ value, it can be seen from \cref{fig:phonon_dispersion_compare_omat}(left) that the phonon dispersion curves computed from the MACE model are well overlapped with the DFT ones, except some deviations near 0 THz around the $X$-high symmetry point. For the worst case of \ce{Na2CuAuCl6}, \cref{fig:phonon_dispersion_compare_omat}(right) shows the MACE underestimates the imaginary phonon frequencies which consequently led to flatter dispersion curve that increases $\mbox{RMSE}(\bm{v}_g)$. In this particular case, the material that is deemed unstable on the DFT-PES became more stabilised on the MACE-PES. 

By further highlighting each point on \cref{fig:harmonic_phonon_summary_plots}(a) with the anharmonic score $\sigma^{(2)}$ computed with DFT for the corresponding HDP structure, a more intriguing trend is revealed, which shows that the accuracy of the phononic properties predicted by the MACE foundation model are strongly correlated with $\sigma^{(2)}$, such that structures with high mechanical stabilities (lower  $\sigma^{(2)}$) exhibit lower RMSEs, and \emph{vice versa}. This is a systematic trend that occurs in all four halide systems investigated, regardless on which datasets the foundation models had been trained  [\cref{fig:landscape-pbe-omat} to \cref{fig:landscape-omat-0-medium}]. This is not a coincidence, as discussed in the Methodology section, that $\sigma^{(2)}$ is also a RMSE-type measurement, but with a fundamental geometric insight that captures the deviation of the shape of the PES from a hyperparabola.  

In \cref{fig:harmonic_phonon_summary_plots}(b) and (c), we show the box plots that provide a more summative view over our benchmark results on the harmonic phonon properties for HDPs. It can be seen that, across the chemical space from fluorides to iodides, the accuracy in predicting the harmonic phonon properties increases as the atomic masses of the halide anions increase. This effect is particularly pronounced in reproducing the phonon eigenfrequencies [\cref{fig:harmonic_phonon_summary_plots}(c)]. As shown in our previous work\cite{yang2022mapping}, the vibrational anharmonicities of HDPs do exhibit a systematic decrease from light to heavier halides [\cref{fig:sigma_landscape_DFT}(a)], hence the chemical trends behind the RMSE values shown in \cref{fig:harmonic_phonon_summary_plots}(b) and (c) is largely consistent with the trend shown in \cref{fig:harmonic_phonon_summary_plots}(a) for chlorides with respect to the variations in $\sigma^{(2)}$. The observed chemical trend is largely unchanged across all four parameterisations of the MACE foundation models, with the \texttt{omat-0-medium} being the best performing model, showing an order of magnitude improvement in $\mbox{RMSE}(\omega^2)$ compared to the worst performing \texttt{matpes-pbe-omat-ft} model. 

On a more fundamental level, the chemical trend observed in RMSEs can be further correlated with the phonon bandwidths (how wide $\omega$ spans, which can be equivalently be reflected from the averaged phonon eigenfrequencies $\langle  \omega\rangle$) for HDPs with different halide anions. As shown in \cref{fig:sigma_landscape_DFT}(b) extracted from our previous work\cite{yang2022mapping}, $\langle\omega\rangle$ increases systematically from iodides to fluorides. This indicates that, for lighter halides, the vibrating ions experience larger restoring forces that originate from a steeper topology of the PES.  From the perspective of training atom-atom force fields\cite{van2016beyond}, it is generally understood that steep or sharp rising parts of the PES are more challenging to be accurately trained, which would require more training data and/or more tailored functional forms to reduce the training complexity. In the domain of fully data-driven MLIPs, the quality of the potential energy model becomes more critically dependent on the breadth of the configuration space covered in the training set. Whilst the \texttt{OMAT}\cite{barroso2024open} dataset already contains rattled atomic structures sampled according to the Boltzmann distribution up to 1000 K, the number of the rattled structures per compound was fixed. Our present findings suggest that, moving forward, a more adaptive scheme in constructing the training sets, particularly applying a weighting scheme to include more rattled structures following the high-frequency phonon modes for systems containing light elements, would be an interesting path to explore for increasing the accuracy of foundational energy models.  

\subsection{Anharmonicity of AIMD-Sampled Configurations from the MACE Foundation Models}
\label{sect:sigma_aimd_traj}

Whilst harmonic phonon properties are often applied first as a key determinant for materials' mechanical stability, in many cases, they are insufficient for fully characterising the finite-temperature phase stabilities of materials. For example, as shown in \cref{fig:phonon_dispersion_compare_omat}, the presence of imaginary phonon frequencies (from calculations performed at 0 K) is often considered as an indicator of mechanical instability. This is a typical feature in many perovskites, however, upon the rise of temperature,  the collective vibrations of ions in the material change the crystal potential that is experienced by the vibrating ions, a physical effect that can be captured by MD simulations. Consequently, the imaginary phonon frequencies become thermally `renormalised' into real ones,\cite{tadano2018first} \emph{i.e.} the material is thermally stabilised at the finite temperature. This shows that MD simulations are essential for fully characterising the finite-temperature phase stabilities of materials, and the capability for MLIPs to generate an ensemble of configurations at a given temperature stably is an important criterion to benchmark the quality of MLIPs.\cite{fu2022forces}

Nevertheless, directly comparing an AIMD trajectory with another one that is independently sampled with a different energy model, in this case, MLIP, is often difficult to come up with good interpretations that can lead to direct and meaningful physical insights into the qualities of MLIPs. Fundamentally, this is because two different energy models correspond to two different PES, and even a small difference in the PES topologies can shift the equilibrium positions, barrier heights and transition states, such that the two MD trajectories may cover completely different configuration spaces. 

To overcome such a complication, in this section, we shall first take the existing AIMD trajectory for each HDP\cite{yang2020mapping} (a total of 1682 valid ones), to recompute the atomic forces for each frame in every trajectory with the MACE foundation model, from which a new $\langle\sigma^{(2)}\rangle_t^{\mbox{\scriptsize{MACE}}}$ metric ($\sigma_{\mbox{\scriptsize{MACE}}}$ for short-handed notation) can be attained, that is to be directly compared with $\langle\sigma^{(2)}\rangle_t^{\mbox{\scriptsize{DFT}}}$ ($\sigma_{\mbox{\scriptsize{DFT}}}$ for short-handed notation). More specifically, for each AIMD trajectory and MACE foundation model that we benchmarked, we compute $\sigma^{\mbox{\scriptsize{MACE}}}$ with two different approaches to obtain the harmonic component of the atomic forces ($F_{i}^{\alpha,(2)}=-\mathbf{\Phi}_{ij}^{\alpha\beta}\mathbf{u}_{i}^{\alpha}$) via the force constant $\mathbf{\Phi}_{ij}^{\alpha\beta}$: (a) $\mathbf{\Phi}_{\mbox{\scriptsize{DFT}}}$-approach, where the atomic forces for each displaced configuration generated from the finite-displacement method\cite{togo2015first} are computed with DFT\cite{yang2022mapping} to construct the force constant $\mathbf{\Phi}$, and (b) $\mathbf{\Phi}_{\mbox{\scriptsize{MACE}}}$-approach, with the atomic forces computed with the MACE foundation models. 

Geometrically, the $\mathbf{\Phi}_{\mbox{\scriptsize{DFT}}}$-approach can be considered as a way of providing a direct measure of the ability of the MACE energy models to exactly reproduce the topologies of the DFT-PES around the local minimum. When  $\sigma_{\mbox{\scriptsize{MACE}}}>\sigma_{\mbox{\scriptsize{DFT}}}$, the MACE model overestimates the total forces, leading to a more anharmonic PES compared to the  DFT baseline, which is the other way around when  $\sigma_{\mbox{\scriptsize{MACE}}}<\sigma_{\mbox{\scriptsize{DFT}}}$. In other words, the discrepancy between $\sigma_{\mbox{\scriptsize{MACE}}}$ and $\sigma_{\mbox{\scriptsize{DFT}}}$ provided an absolute measure on the errors in predicting the total atomic forces. When $\sigma_{\mbox{\scriptsize{MACE}}}=\sigma_{\mbox{\scriptsize{DFT}}}$, the DFT energy landscape can be fully reconstructed by the MACE energy models over all $\{\mathbf{u}\}$.  For the $\mathbf{\Phi}_{\mbox{\scriptsize{MACE}}}$-approach, $\sigma_{\mbox{\scriptsize{MACE}}}$ takes no reference to the DFT-energy landscape, thus it reflects the degree of anharmonicity of the MACE-energy landscape itself. When  $\sigma_{\mbox{\scriptsize{MACE}}}=\sigma_{\mbox{\scriptsize{DFT}}}$, it means that the relative contributions from the (an)harmonic force components to the total atomic forces are the same between the MACE and DFT energy models, and the MACE and DFT-PES differ from each other globally by some constant multiplicative factor. 

\begin{figure}[tb]
    \centering
    \includegraphics[width=\linewidth]{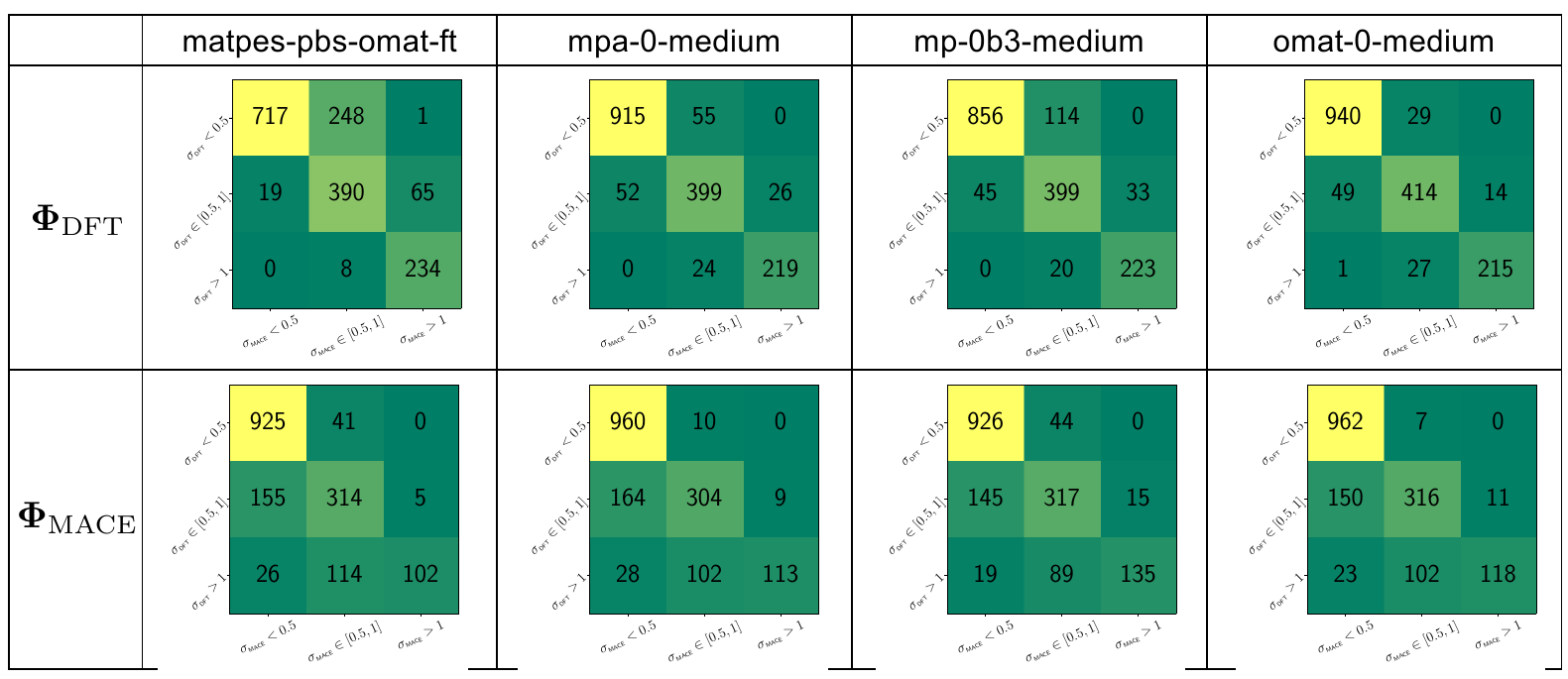}
    \caption{Table of confusion matrices showing how well the MACE models (across the columns) reproduce the vibrational anharmoncity of DHPs computed from DFT (across the rows), which is represented as the number of DHPs that fall into each category. Top (bottom) row presents  the anharmonicity scores evaluated using force constant matrix computed from DFT (corresponding MACE models shown on the top row).} 
\label{fig:confusion_matrix_aimd_traj}
\end{figure}

Physically, comparing $\sigma_{\mbox{\scriptsize{MACE}}}$ with $\sigma_{\mbox{\scriptsize{DFT}}}$ provides the key indication on whether the degrees of finite-temperature dynamic stabilities of materials predicted by the MACE foundation models agree with those predicted by the DFT. In particular, the $\mathbf{\Phi}_{\mbox{\scriptsize{MACE}}}$-approach provides a looser criterion in making this judgement as it only requires the relative contributions of the anharmonic forces to the total one being the same as predicted from MACE models and DFT, which may benefit from error cancellations in subtracting $\mathbf{\Phi}_{ij}^{\alpha\beta}\mathbf{u}_{i}^{\alpha}$ from the total atomic force $F_i^\alpha$, when both terms are predicted from the MACE models. In contrast, the   $\mathbf{\Phi}_{\mbox{\scriptsize{DFT}}}$-approach is more strict, which would require the total atomic force predicted from the MACE-models to closely match those computed from DFT. These information are presented with the confusion matrices shown in \cref{fig:confusion_matrix_aimd_traj}. In each confusion matrix, the dynamic stabilities are characterised into three categories\cite{yang2022atlas}: (a) $\sigma\in(0,0.5)$, corresponding to highly stable structures with weak vibrational anharmonicity that is predominantly contributed by three-phonon scatterings (encapsulated by the third-order force constant $\mathbf{\Phi}_{ijk}^{\alpha\beta\gamma}$). (b)  $\sigma\in[0.5,1]$, meaning the phase is still stable at the simulated temperature $T$ but with stronger vibrational anharmonicity that is dominated by the force constants from fourth-order and above. And finally, (c) $\sigma>1$, meaning the phase is unstable at $T$.

Results from \cref{fig:confusion_matrix_aimd_traj} show that, the confusion matrices are dominated by the diagonal elements, meaning that the consensuses in predicting the phase stabilities at 300 K between the MACE foundation models and DFT are generally acceptable across a wide range of stability regimes, supporting the argument that MACE foundation models is useful for fast pre-screening filter for unstable materials\cite{batatia2023foundation}. 

In the $\mathbf{\Phi}_{\mbox{\scriptsize{DFT}}}$-approach, one sees a clear trend of increasing number of correctly predicting materials' dynamic stabilities from the \texttt{matpes-pbs-omat-ft} to the \texttt{omat-0-medium} model, which is in line with the observations from \cref{sect:harmonic_phonons}. The likelihood for the MACE models to overly stabilising (destabilising) the DFT-predicted unstable (stable) materials, \emph{i.e.} $\sigma_{\mbox{\scriptsize{MACE}}}>1$ for $\sigma_{\mbox{\scriptsize{DFT}}}<1$ or \emph{vice versa} are generally low. Except the \texttt{matpes-pbs-omat-ft}  and \texttt{mp-0b3-medium} models, for which we see a significant number of weakly anharmonic HDPs being classified as strongly anharmonic by the MACE models. 

The $\mathbf{\Phi}_{\mbox{\scriptsize{MACE}}}$-approach reveals a different outcome. In this case, the number of HDPs that have their dynamical stabilities being correctly identified remain almost unchanged across different MACE models.  As discussed above, this means that, across a large part of the chemical space, the relative anharmonic contributions to the overall topologies of the PES remain invariant from DFT to the different MACE models. However, we also observed from \cref{fig:confusion_matrix_aimd_traj} that when the $\mathbf{\Phi}_{\mbox{\scriptsize{MACE}}}$ is used to extract the harmonic components of the atomic forces, the number HDPs being placed in the lower off-diagonal parts of the confusion matrices increased significantly, meaning when the MACE model is used solely to compute $\sigma$, it tends to over-stabilise the highly anharmonic and unstable HDPs (see \cref{fig:sigma_trajectory_examples}(d) for an example). This observation can be better reflected when we plot the distributions of  $\langle \sigma^{(2)}\rangle_t^{\mbox{\scriptsize{MACE}}}-\langle \sigma^{(2)}\rangle_t^{\mbox{\scriptsize{DFT}}}$ in \cref{fig:aimd_traj_mace_phi_benchmark}), which show strong tailing in the negative part. As discussed in \cref{sect:harmonic_phonons}, this reflects the poorer generalisability of the MACE foundational models in capturing the anharmonic features of the PES, particularly for materials with low dynamical stabilities.

\subsection{Anharmonicity of HDPs Computed Solely from the \texttt{omat-0-medium} Model}

In the practical applications where MLIP is used to determine the dynamic stabilities of materials, both the harmonic force constants and the finite-temperature MD sampling would have been conducted with the same MLIP, with little or no reference to prior DFT results. Hence, in this section, we shall present some further results and analysis on determining the vibrational anharmonicity of HDPs solely based on the \texttt{omat-0-medium}, which was shown to be the best performing model from the previous sections. 

Computationally, the MD samplings using the MACE foundation model, dubbed as MACE-MD, are carried out as follows. For each HDP, we randomly selected 2 frames from the previously sampled AIMD trajectory as the starting points to perform 2 independent MACE-MD samplings. Such a choice of the starting points for MACE-MD imposes a weak constraint that the configurational space that is sampled by the MACE-MD should have some overlap with the configurational space sampled from AIMD, at least in the initial stage of the MACE-MD sampling. Each MACE-MD simulation was ran for 2 fs at 1 ps time step using the MD engine from the Atomic Simulation Environment\cite{ase-paper}. Same as our previous work\cite{yang2022mapping}, the Andersen thermostat\cite{andersen1980molecular} with a collision probability of 0.5 was employed to maintain the simulation temperature at the target value of 300 K. The corresponding $\langle\sigma^{(2)}\rangle_t^{\mbox{\scriptsize{MACE}}}$ was averaged over all 4000 MACE-MD frames using the force constant $\bm{\Phi}_{\mbox{\scriptsize{MACE}}}$ computed with the same \texttt{omat-0-medium} model to extract the harmonic components of the atomic forces.

\begin{figure}[tbh]
    \centering
    \includegraphics[width=0.95\linewidth]{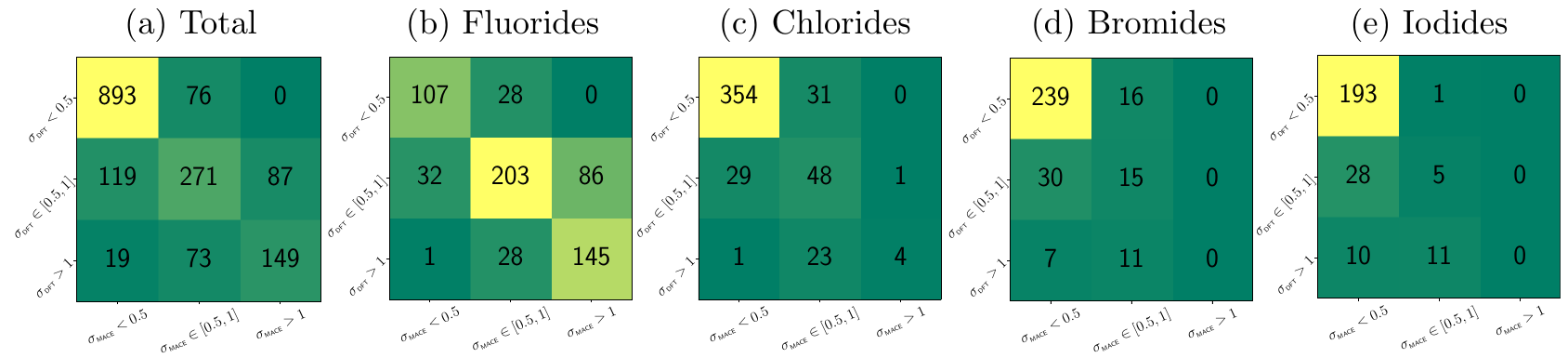}
    \caption{Confusion matrices showing the reproducibility of the DFT-computed vibrational anharmonicity for DHPs using the MACE model. Here, the \texttt{omat-0-medium} model is used for both computing the harmonic force constants, as well as the molecular dynamics samplings.}
    \label{fig:confusion_matrix_mlff_traj}
\end{figure}

\cref{fig:confusion_matrix_mlff_traj} presents the confusion matrices that compare the numbers of HDPs that share the same/different classifications of their dynamic stabilities as solely determined from either the DFT or the \texttt{omat-0-medium} foundation model. Similar to the results shown in \cref{fig:confusion_matrix_aimd_traj}, the confusion matrix  [\cref{fig:confusion_matrix_mlff_traj}(a)] is still dominated by the diagonal elements, meaning that the performance in determining the room-temperature dynamic stabilities using the \texttt{omat-0-medium} foundation model alone is generally acceptable. Counting the numbers in the lower diagonal part of the confusion matrix, we found 38 \% chance of categorising HDPs with low stabilities to be more stable ones. In contrast,  the upper diagonal part of the confusion matrix leads to only 6 \% chance of misplacing stable materials to be less stable, which predominantly comes from the fluorides [\cref{fig:confusion_matrix_mlff_traj}(b)]. This shows that the \texttt{omat-0-medium}  model leads to more false positive cases than false negative one. This means that the chance of missing stable materials is lower, compared to including more unstable materials,  when it comes to (pre-)screening dynamically stable materials using the \texttt{omat-0-medium}  model, whereby more accurate models (such as DFT) can be used subsequently to further filter out the false positive results.

\begin{table}[tb]
\begin{tabularx}{\linewidth}{*{2}{>{\        \arraybackslash}X}}
\begin{tabular}[b]{p{2cm}p{1.25cm}p{1.25cm}p{3cm}}
    \hline\hline
        Compound &  $\sigma_{\mbox{\scriptsize{DFT}}}$ & $\sigma_{\mbox{\scriptsize{MACE}}}$ & $\left\vert \sigma_{\mbox{\scriptsize{DFT}}}-\sigma_{\mbox{\scriptsize{MACE}}}\right\vert$\\
    \hline
     \ce{Rb3AlF6}    &  $<0.5$ & $<0.5$ & 0.000375 \\
     \ce{Rb2NiAgF6}    & $<0.5$ & $[0.5,1]$ & 0.400 \\
     \ce{Cs2NaRuF6}    & $[0.5,1]$ & $<0.5$  & 0.663\\
     \ce{K2InAgF6}    & $[0.5,1]$ & $[0.5,1]$  & 0.00102  \\
     \ce{K2RbSbF6}  & $[0.5,1]$ & $>0.5$ & 0.440 \\
    \hline\hline
    \end{tabular}
    &
    \includegraphics[scale=0.6]{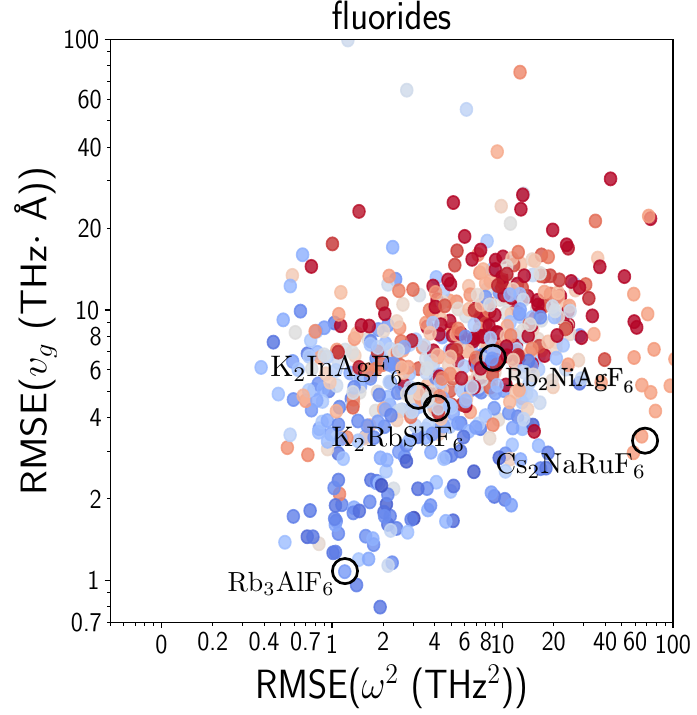}\\   \caption{Selected fluoride HDPs, the molecular dynamics trajectories of which  sampled from AIMD and MACE-MD, will be compared with the SOAP-REMatch kernel.}
    \label{tab:kpca_compounds}
    &
    \captionof{figure}{Plot of the RMSE in predicting the phonon eigenfrequencies and group velocities with respect to the DFT results using the \texttt{omat-0-medium} model, with the locations of the five compounds selected in \cref{tab:kpca_compounds} highlighted on the plot. }
    \label{fig:compound_map}
\end{tabularx}
\end{table}

\begin{figure}[tb]
    \centering
    \includegraphics[width=\linewidth]{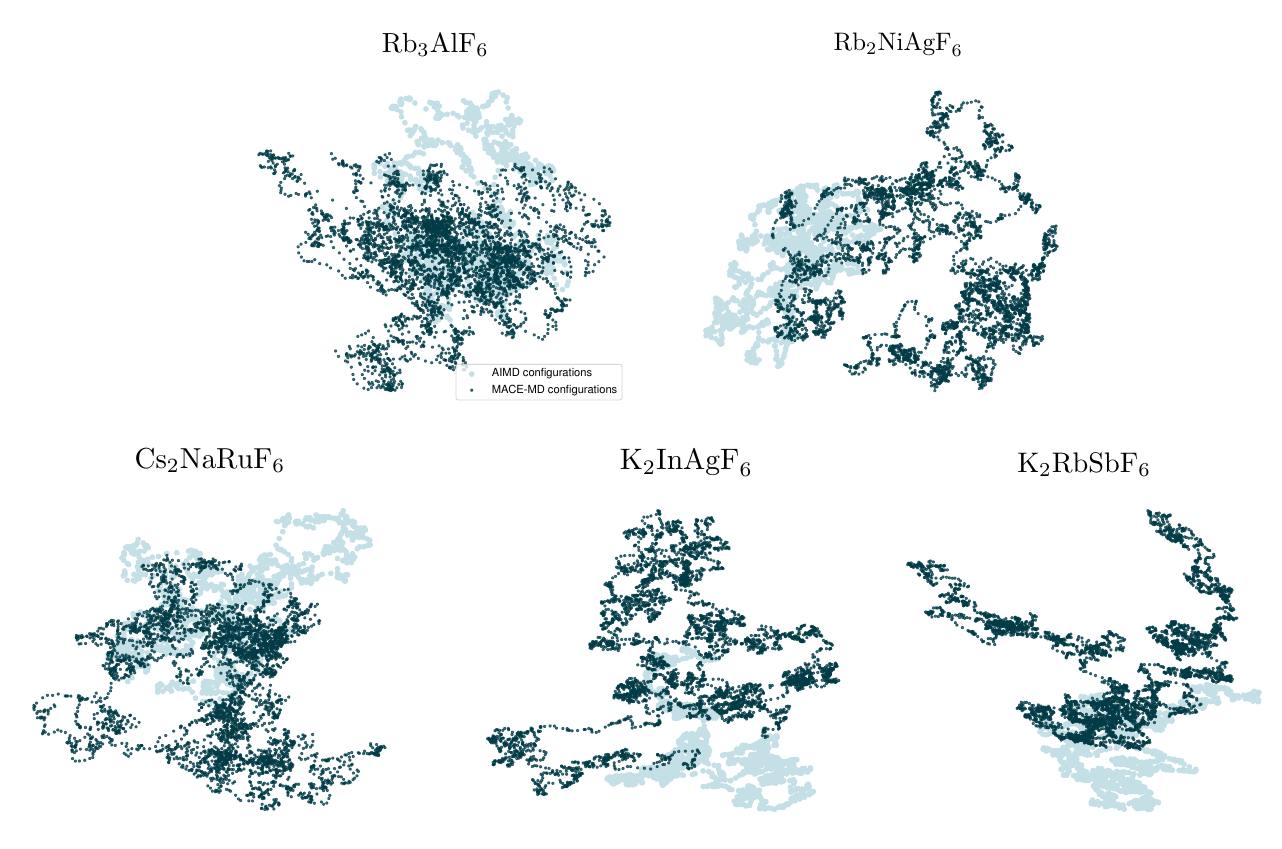}
    \caption{KPCA maps for the five fluoride HDPs listed in \cref{tab:kpca_compounds} that compare the configuration spaces sampled by AIMD\cite{yang2022mapping}  and MACE-MD with the \texttt{omat-0-medium} model. Each point on the map corresponds to a configuration in the MD trajectory.}
    \label{fig:kpca_maps}
\end{figure}

As mentioned in \cref{sect:sigma_aimd_traj}, comparing two energy models using results from MD simulations may introduce bias because the subtle differences in the PES topologies underpinned by the two energy models may cause MD simulations to sample two distinctly different configurational spaces that intrinsically possess different properties. To assess the extent of this bias that could have contributed to the results that are presented in \cref{fig:confusion_matrix_mlff_traj}, we have selected five extreme cases among the fluoride compounds (\cref{tab:kpca_compounds}) and performed unsupervised machine learning to compare the similarities in the configurational spaces that are sampled by the AIMD and MACE-MD [see \cref{sect:method} for details]. Results from such analysis (\cref{fig:kpca_maps}) show that, first of all, the way we selected the initial structures for running the MACE-MD simulations did mitigate some of the bias by enforcing the configurational spaces sampled by the two different energy models to (at least partially) overlap with each other. System that exhibits the largest overlap in the configuration spaces sampled by the two energy models is \ce{Rb3AlF6}, of which the computed $\sigma_{\mbox{\scriptsize{MACE}}}$ is literally identical to $\sigma_{\mbox{\scriptsize{DFT}}}$ (\cref{tab:kpca_compounds}). In this case, we can consider the DFT-PES around the local minimum for the cubic \ce{Rb3AlF6} has been well reproduced by the \texttt{omat-0-medium} model. On the contrary, \ce{K2RbSbF6} represents the other extreme case where the configuration space sampled by the MACE-MD diverges quite significantly from the one sampled with AIMD. The other three compounds listed in \cref{tab:kpca_compounds} are somewhere between these two extreme cases, as revealed in \cref{fig:kpca_maps}.

    

To check that the above observations are not necessarily biased by the longer trajectories that are sampled by the numerically more efficient MLIP, we performed extra simulations which extended the original AIMD trajectories to 4 ps in length, and re-performed the same KPCA analysis. With longer AIMD trajectory, \cref{fig:kpca_maps_long_DFT} shows the divergence between AIMD and MACE-MD trajectories for \ce{K2RbSbF6} has reduced, but for \ce{Rb2NiAgF6}, the divergence between the two trajectories increased. Taking into account the stochasticity of MD simulations when interpreting the KPCA maps that are shown in \cref{fig:kpca_maps} and \cref{fig:kpca_maps_long_DFT_sigma}, we can say that the topologies of the PES underpinned by DFT and the \texttt{omat-0-medium} model for most materials under the current investigation should be very similar, rendering sufficient similarities in the configuration spaces that are sampled from these two models. As such, dissimilarities in the sampled configuration spaces are not believed to be strongly contributing to the discrepancies in $\sigma_{\mbox{\scriptsize{DFT}}}$ and $\sigma_{\mbox{\scriptsize{MACE}}}$. 

By further colouring each point on the KPCA maps with the anharmonicity score for the corresponding MD snapshot (\cref{fig:kpca_maps_sigma} and \cref{fig:kpca_maps_long_DFT_sigma}), it becomes clear that when $\left\vert\sigma_{\mbox{\scriptsize{DFT}}}-\sigma_{\mbox{\scriptsize{MACE}}}\right\vert$ is large, it can be generally attributed to a systematic error in which $\sigma_{\mbox{\scriptsize{MACE}}}$ computed for the entire MACE-MD trajectory is collectively and significantly different from $\sigma_{\mbox{\scriptsize{DFT}}}$ even in the regions of the configurational space where the overlap between those sampled from AIMD and MACE-MD is significant. This suggests that the error in computing $\sigma$ must be inherited from the error in computing the Hessian matrix $\mathbf{\Phi}$, which is indeed supported by \cref{fig:compound_map} showing that low (high) errors in predicting the phonon group velocities and frequencies generally translate to low (high) discrepancies between predicted values of $\sigma_{\mbox{\scriptsize{MACE}}}$ and $\sigma_{\mbox{\scriptsize{DFT}}}$. The reason for this, is that, understandably, just as the higher accuracy that is required in calculating the atomic forces for determining the phononic properties with DFT, even small errors in predicting the atomic forces with the foundational models could translate into large notable differences in the phonon dispersion relationship, as the errors are amplified in the calculations of the derivatives of forces with respect to the atomic positions.  

\section{Conclusions and Outlooks}

Using our own unique database that has characterised the degree of vibrational anharmonicity and room-temperature dynamic stabilities of $\sim2000$ halide double perovskites, which includes both the harmonic phonon and 300 K-MD simulation data computed at DFT level of theory, in this work, we have systematically benchmarked four latest variants of the MACE foundation models for inorganic solids, on their capabilities and accuracies in determining materials' dynamic stabilities. This is an important application scenario for the foundation models in computational material discoveries, where phase stabilities predicate all subsequent endeavours of discovering new exotic physical and chemical applications of new materials.  Out of the four variants of the MACE foundation models, it is found that the \texttt{omat-0-medium} model performs the best in reproducing both the 0 K-harmonic phonon properties, as well as the room-temperature dynamic stabilities of HDPs that were determined from DFT simulations. Mathematically, the arharmonicity score shares a highly similar form as the standard deviations that are employed to measure the accuracy of the MLIPs, thus it is reasonable to observe that the errors in predicting the harmonic phonon properties using the MACE foundation models showed strong correlation with the structures' anharmonicity scores, whereby weakly (strongly) anharmonic materials exhibit higher (lower) accuracies in such predictions. This suggests that including more data, such as high-temperature MD trajectories, metastable materials, or even hypothetical materials that may be unstable, is important in further developing and/or fine-tuning foundation models to achieve broad applicability and better generalisability. 

Based on the above primary findings, we further extended our benchmark by computing the anharmonicity scores for HDPs with both the harmonic force constants and MD samplings solely based on the \texttt{omat-0-medium} model. It is found that the dynamic stabilities determined using such an approach correlate well with the DFT results, suggesting that the \texttt{omat-0-medium} model is suitable for accelerating the screening of dynamic materials stabilities for materials discoveries. In more careful examinations of the HDP systems, on which the \texttt{omat-0-medium} model performed well (pooly) in reproducing the anharmoncity scores as determined by DFT, we think  it is reasonable to believe that the topologies of the DFT PES are generally well reproduced by the MACE foundation model, whereby considerable overlaps in the configuration spaces sampled by the two approaches can be observed. The (large) discrepancies between the foundation-model- and DFT-predicted anharmonicity scores can be predominantly attributed to the amplification of the errors in predicting the atomic forces with the foundation model when calculating the Hessian matrices.  This also shows the possible need of including Hessians in training materials' foundation models, to promise their applications in scenarios where high numerical precision is a must in atomistic materials' modelling.

We hope that the findings presented in this study have provided interesting and useful insights to facilitate the ongoing developments and fine-tunings of materials foundation models. For instance, proposing metrics such as the anharmonic score is particularly interesting, which it not only can be used for quantifying the model quality, but also be interpreted based on materials' properties to provide more physical insights in understanding the model performances. We envisage that the continuous evolution of the foundation models will further advance the important statistical physics tools in configurational space sampling, particularly in tackling the challenge of meeting the ergodic condition, which bears implications in computing and understanding a wide range of physical and chemical properties of functional materials, such as thermal expansions, lattice thermal conductivities, catalytic activities under realistic (\emph{e.g.} solvated) environments, and many others.

{\small
\singlespacing
\bibliography{ref.bib}
}

\newpage
 
\setcounter{section}{0}
\setcounter{figure}{0}

\renewcommand{\thesection}{S\arabic{section}}
\renewcommand{\thesubsection}{S\arabic{section}.\arabic{subsection}}

\renewcommand{\thefigure}{S\arabic{figure}}

\section*{Appendix}

\section{Database Overview}
\subsection{Data availability}
\label{sect:datalink}
All DFT data for our HDP database is stored in the Harvard Dataverse, which can be freely accessed from the following links:
\begin{enumerate}
    \item Fluoride HDPs: \url{https://doi.org/10.7910/DVN/WBOXPG}
    \item Chloride HDPs: \url{https://doi.org/10.7910/DVN/JGODBE}
    \item Bromide HDPs: \url{https://doi.org/10.7910/DVN/RIMZ2F}
    \item Iodide HDPs: \url{https://doi.org/10.7910/DVN/ATZEFE}
\end{enumerate}

\subsection{Vibrational anharmonicity landscape at DFT level of theory}
\begin{figure}[htb]
    \centering
    \includegraphics[width=\linewidth]{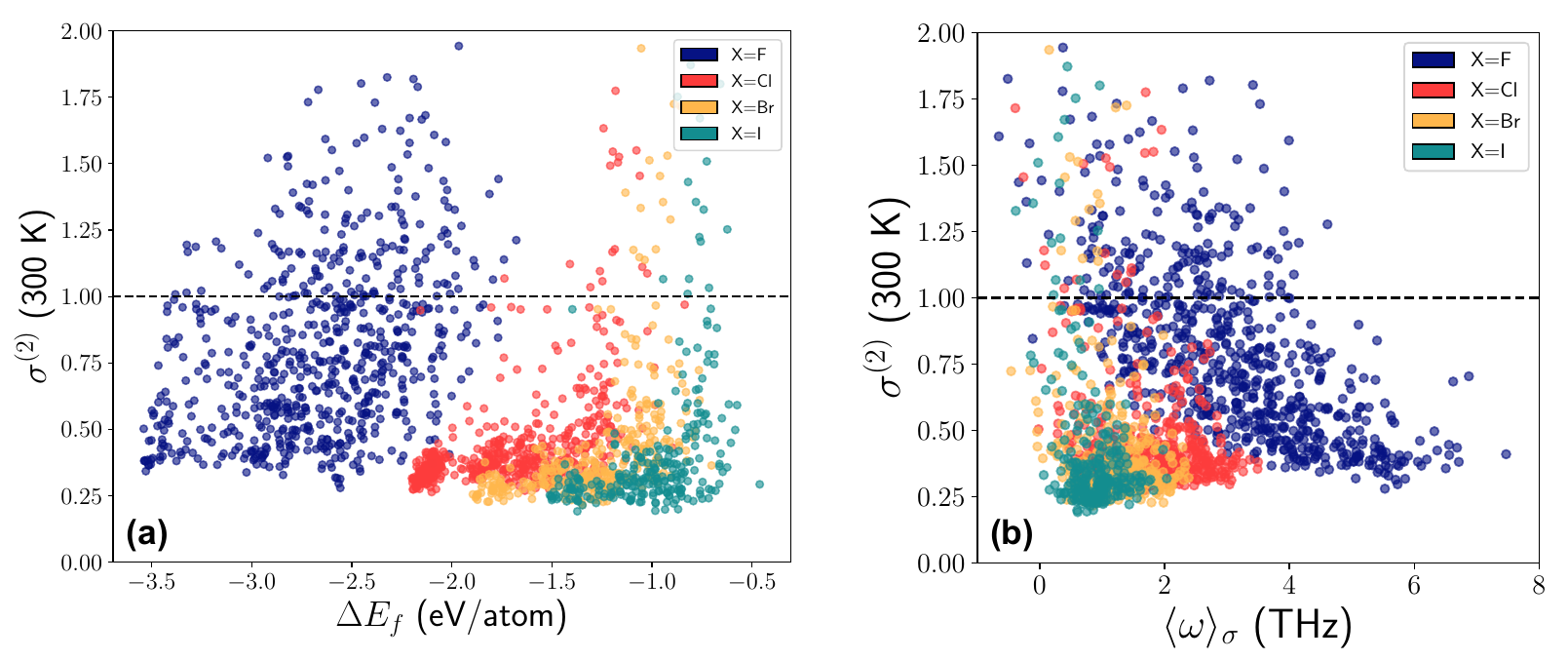}
    \caption{Landscape of room-temperature vibrational anharmonicities as measured by $\sigma^{(2)}$ as a function of formation energies for HDPs, in which data results for HDPs with different halogen anions are separately colour-coded.  (a) $\sigma^{(2)}$ plotted as a function of the formation energies $\Delta  E_f$. (b)  $\sigma^{(2)}$ plotted as a function of anharmonicity-weighted-averaged phonon frequency for each HDP, defined as $\langle\omega\rangle_\sigma=\sum_{\bm{q},n}\omega(\bm{q},n)\sigma^{(2)}(\bm{q},n)/\sum_{\bm{q},n}\sigma^{(2)}(\bm{q},n)$, in which the phonon-mode-resolved anharmonicity score $\sigma^{(2)}(\bm{q},n)$  was computed using the same definition as \cref{eq:anharmonic_score} except all the atomic forces are projected onto individual phonon eigenvectors $\bm{u}(\bm{q},n)$  [Reproduced from Yang \emph{et al.}\cite{yang2022mapping}].}
    \label{fig:sigma_landscape_DFT}
\end{figure}

\newpage
\section{Exemplary Phonon Dispersion Curves}
\begin{figure}[htb]
    \centering
    \includegraphics[width=\linewidth]{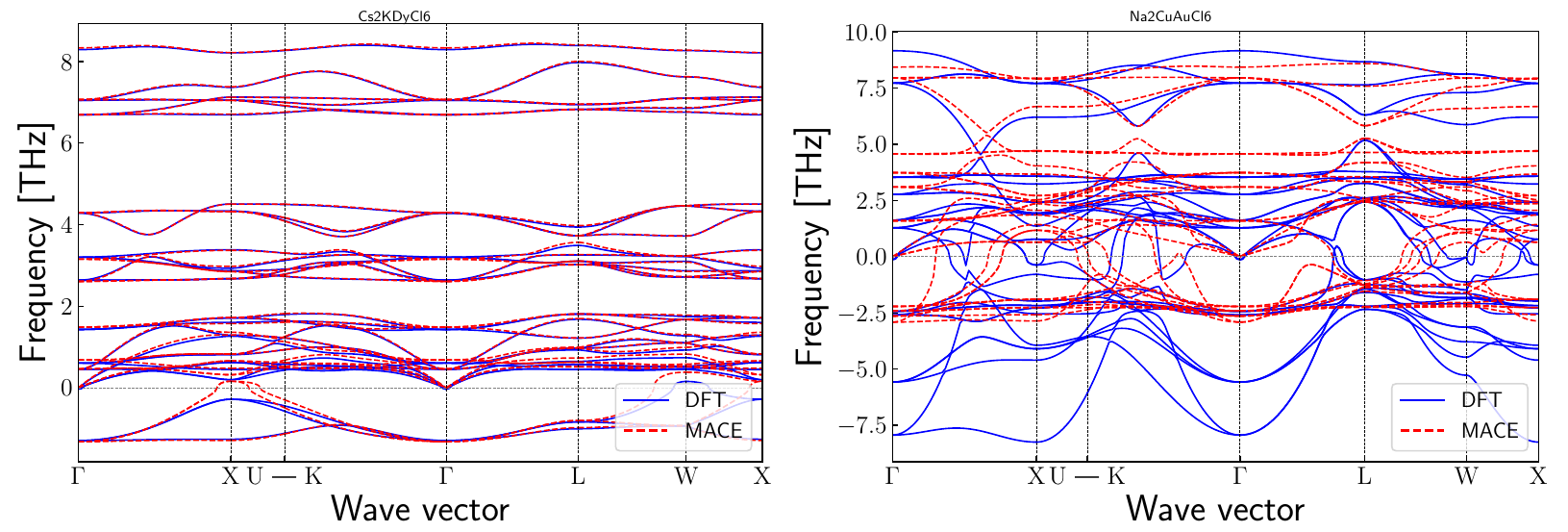}
    \caption{Comparisons of the phonon dispersion relationship computed with DFT and \texttt{omat-0-medium} foundation model. The results shown above correspond to the compounds of (Left) \ce{Cs2KDyCl6} and (Right) \ce{Na2CuAuCl6}, which are the best and worst performing compounds for predicting harmonic phonon properties with the \texttt{omat-0-medium} model, respectively. }
    \label{fig:phonon_dispersion_compare_omat}
\end{figure}

\section{Accuracies of Predicting Harmonic Phonon Properties - Breakdown Analysis in Different Chemical Spaces}

\begin{figure}[h]
    \centering
    \includegraphics[width=\linewidth]{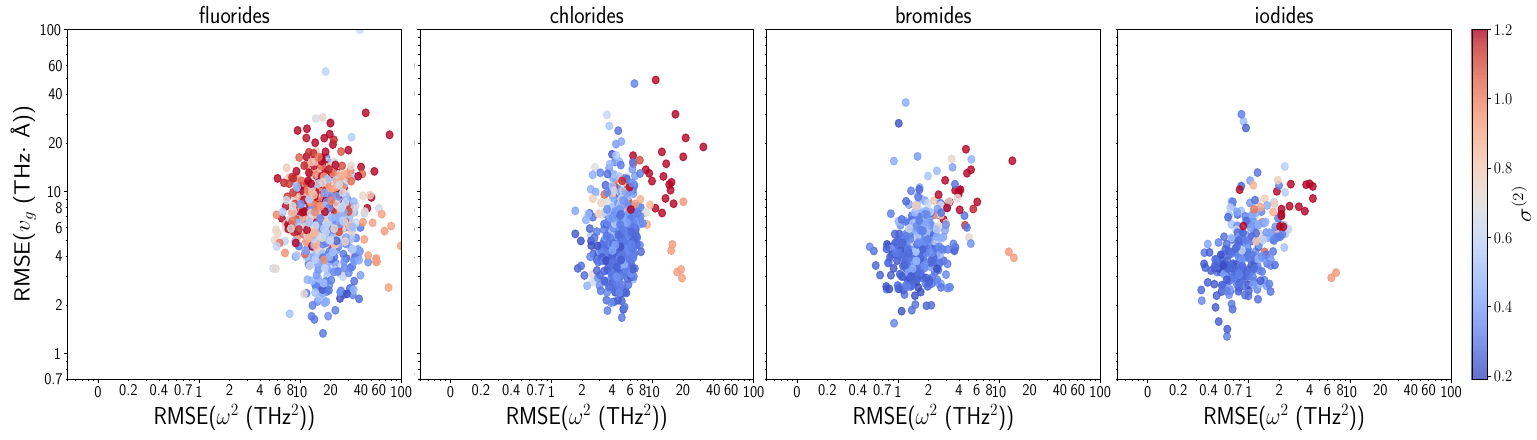}
    \caption{Scatter plots showing the correlations between the root-mean-squared-errors in predicting the phonon eigenfrequencies and group velocities using the MACE foundation model with respect to the DFT results.  Results for HDPs with different halide anions are presented in separate subplots to better highlight the chemical trend. Each point in the plots are colour-coded according to their anharmonicity scores $\sigma^{(2)}$ obtained from DFT calculations\cite{yang2022mapping}. For this set of results, the \texttt{matpbs-pbe-omat-ft} model was used.}
    \label{fig:landscape-pbe-omat}
\end{figure}

\begin{figure}[h]
    \centering
    \includegraphics[width=\linewidth]{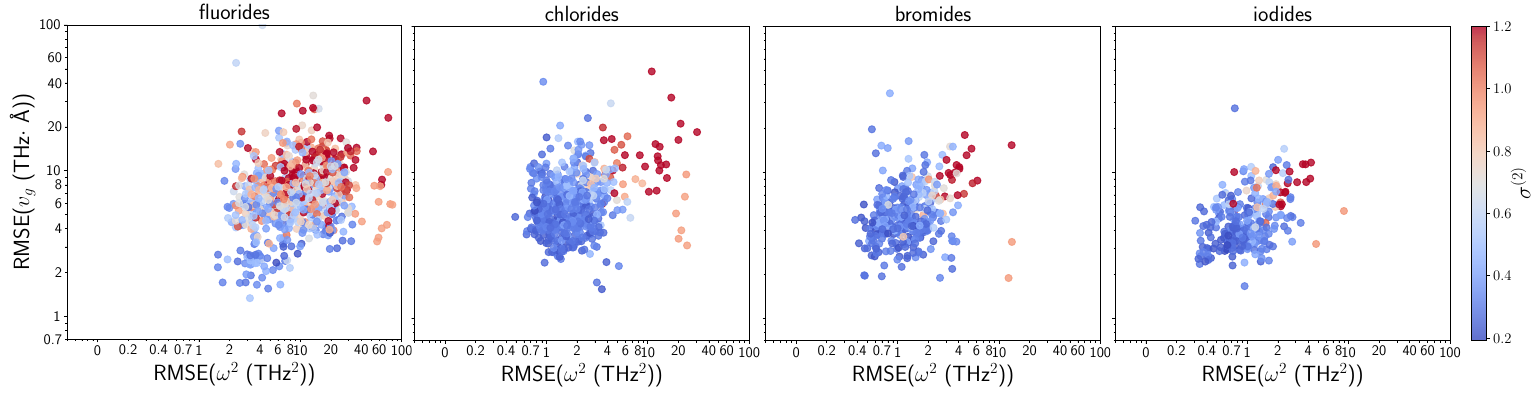}
    \caption{Same as \cref{fig:landscape-pbe-omat} with results obtained using the \texttt{mpa-0-medium} model.}
    \label{fig:landscape-mpa-0-medium}
\end{figure}

\begin{figure}[h]
    \centering
    \includegraphics[width=\linewidth]{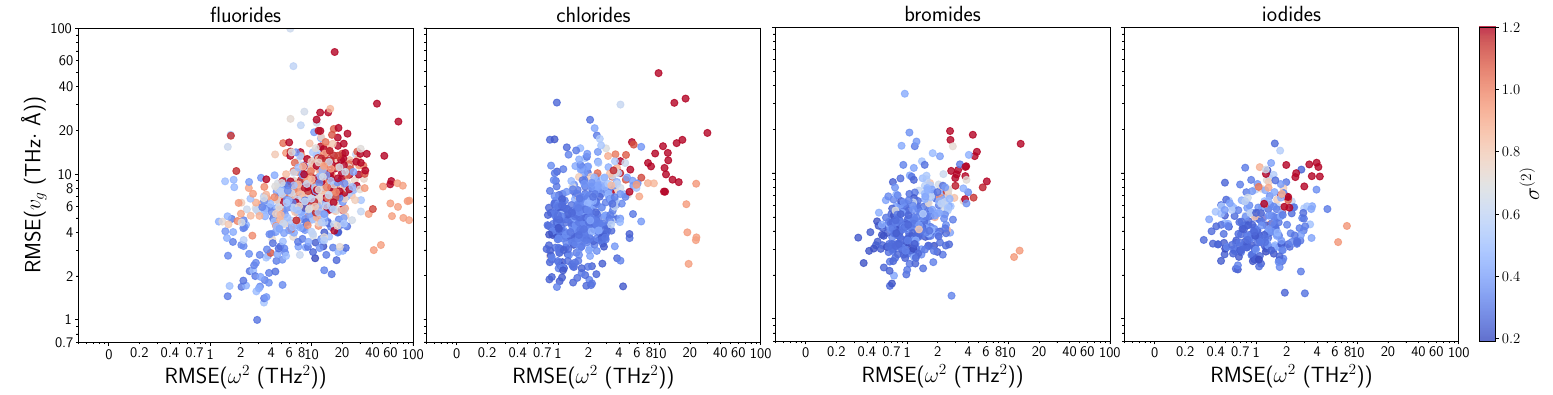}
    \caption{Same as \cref{fig:landscape-pbe-omat} with results obtained using the \texttt{mp-0b3-medium} model.}
    \label{fig:landscape-mp-0b3-medium}
\end{figure}

\begin{figure}[h]
    \centering
    \includegraphics[width=\linewidth]{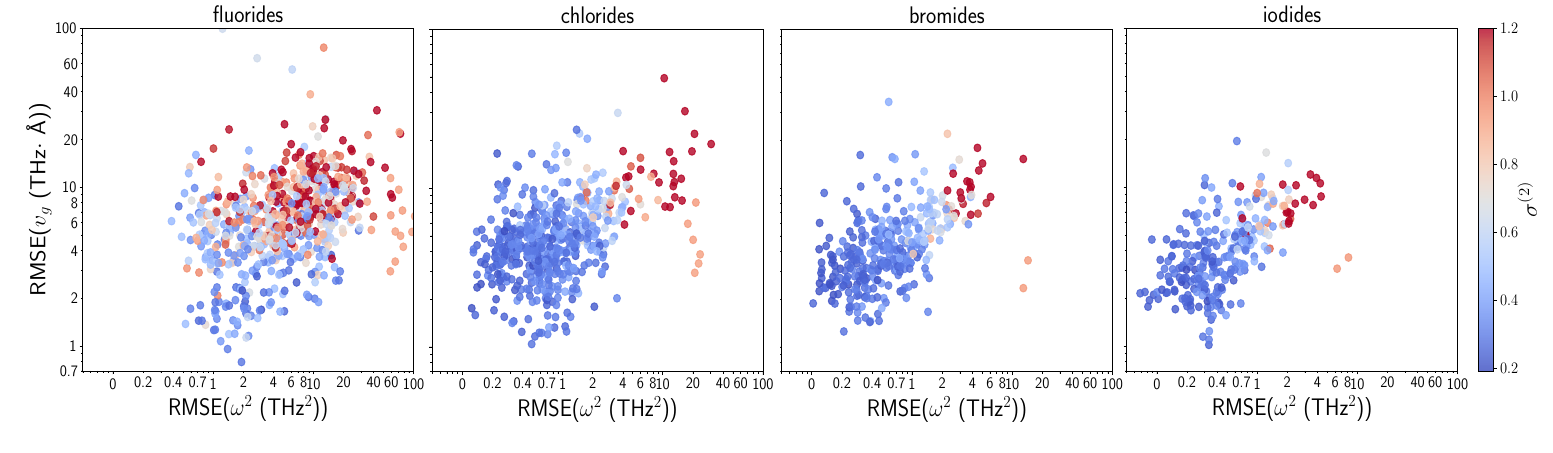}
    \caption{Same as \cref{fig:landscape-pbe-omat} with results obtained using the \texttt{omat-0-medium} model.}
    \label{fig:landscape-omat-0-medium}
\end{figure}

\clearpage

\section{Accuracies of Predicting the Room-Temperature Vibrational Anharmonicity}

\begin{figure}[htb]
    \centering
    \includegraphics[width=0.7\linewidth]{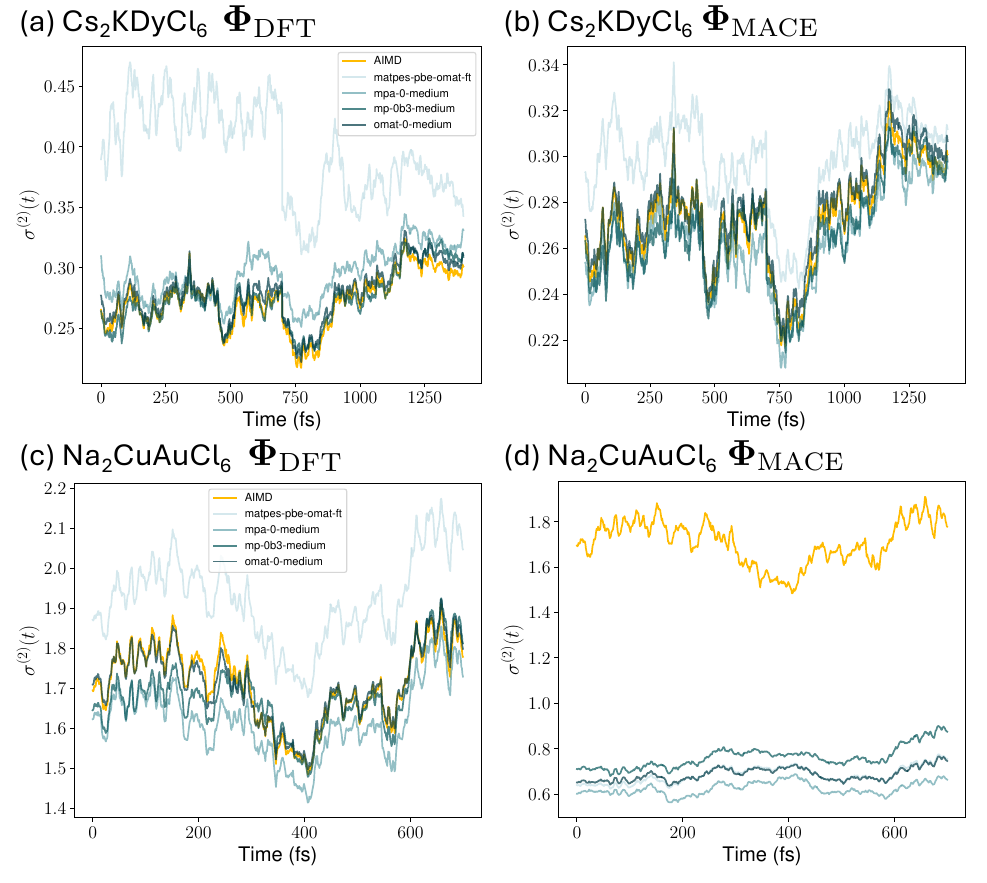}
    \caption{Examples of the $\sigma^{(2)}(t)$ trajectories for two exemplary chloride HDPs [\cref{fig:harmonic_phonon_summary_plots}(a)]. The anharmonic scores  $\sigma^{(2)}$ are determined for the configurations that were previously sampled from AIMD\cite{yang2020mapping}, with the atomic forces for each trajectory frame recomputed by different MACE foundation models. On the left panel, we compare the results whereby the harmonic components of the atomic forces were determined based on DFT-derived force constants ($\mathbf{\Phi}_{\mbox{\scriptsize{DFT}}}$), whereas the right panel shows the case for $\mathbf{\Phi}_{\mbox{\scriptsize{MACE}}}$, in which the force constants were also recomputed using the corresponding MACE foundation models.}
    \label{fig:sigma_trajectory_examples}
\end{figure}

\begin{figure}[htb]
    \centering
    \includegraphics[width=\linewidth]{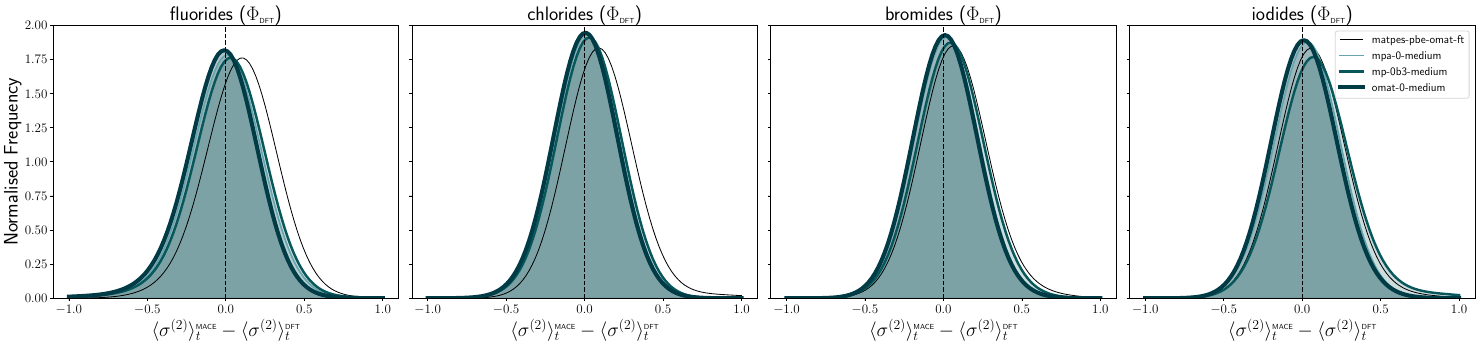}
    \caption{Statistical distributions on the differences in the trajectory-averaged anharmonic scores computed from MACE foundation models and DFT ($\langle \sigma^{(2)}\rangle_t^{\mbox{\scriptsize{MACE}}}-\langle \sigma^{(2)}\rangle_t^{\mbox{\scriptsize{DFT}}}$), both evaluated on the AIMD trajectories. The harmonic force constant matrix calculated from DFT $(\Phi_{\mbox{\scriptsize{DFT}}})$ are used for determining the harmonic component of the atomic forces. Data for DHPs with different halide anions are shown separately. For details, see \cref{sect:sigma_aimd_traj}.}
\label{fig:aimd_traj_dft_phi_benchmark}
\end{figure}

\begin{figure}[htb]
    \centering
    \includegraphics[width=\linewidth]{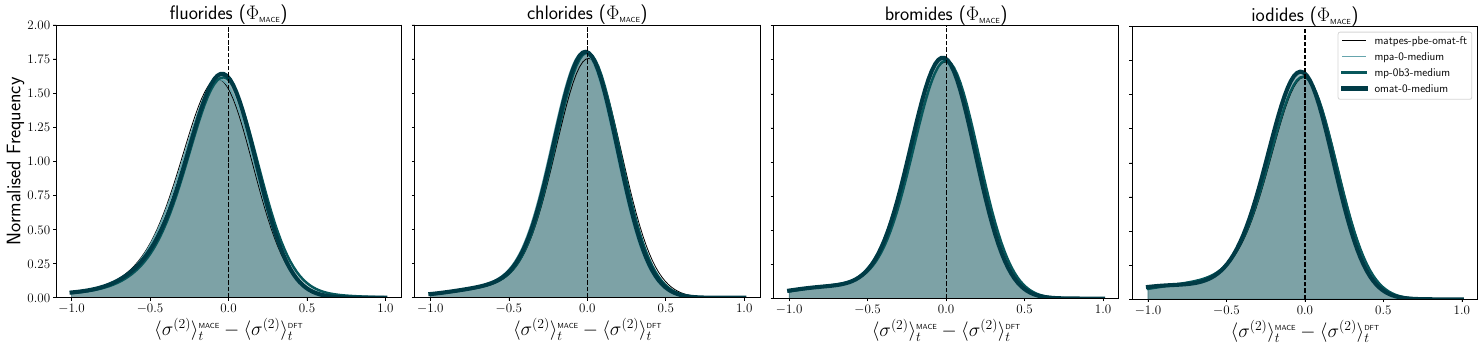}
    \caption{Same as \cref{fig:aimd_traj_dft_phi_benchmark}, except the force constant matrix $(\Phi_{\mbox{\scriptsize{MACE}}})$ are computed from the finite-difference approach using the same MACE foundation model for evaluating the total atomic forces on AIMD trajectory frames.}
\label{fig:aimd_traj_mace_phi_benchmark}
\end{figure}

\clearpage
\section{Additional Sketch Map Analysis}

\begin{figure}[htb]
    \centering
    \includegraphics[width=\linewidth]{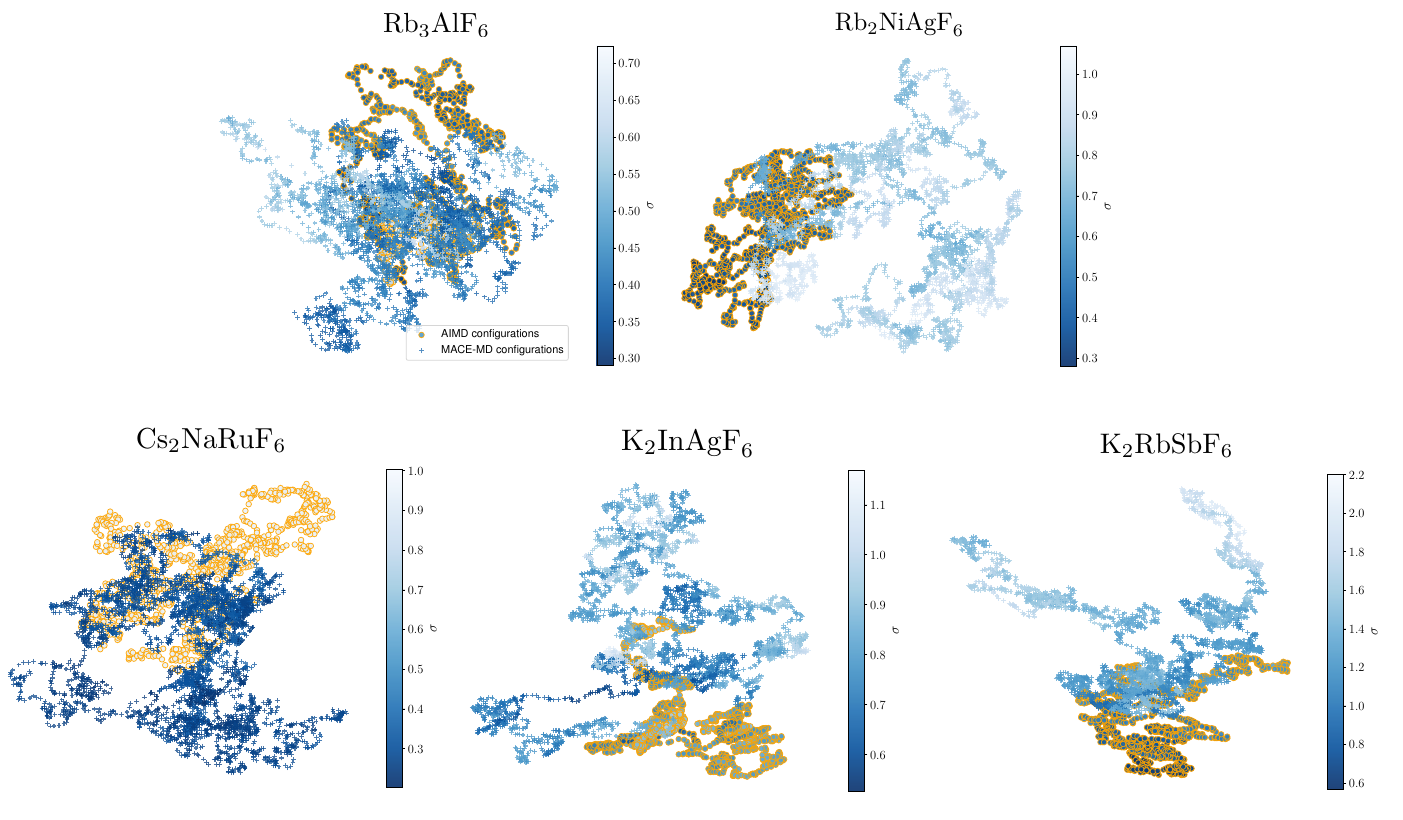}
    \caption{KPCA maps that compare the configurational space sampled by  AIMD\cite{yang2022mapping} and MACE-MD with the \texttt{omat-0-medium} model. Each configuration (point on the KPCA map) is further colour-coded with its anharmonicity score $\sigma^{(2)}$. The harmonic force components necessary for computing $\sigma^{(2)}$ were determined from the force constants that are computed with the same energy model as for the MD simulations.}
    \label{fig:kpca_maps_sigma}
\end{figure}

\begin{figure}[htb]
    \centering
    \includegraphics[width=\linewidth]{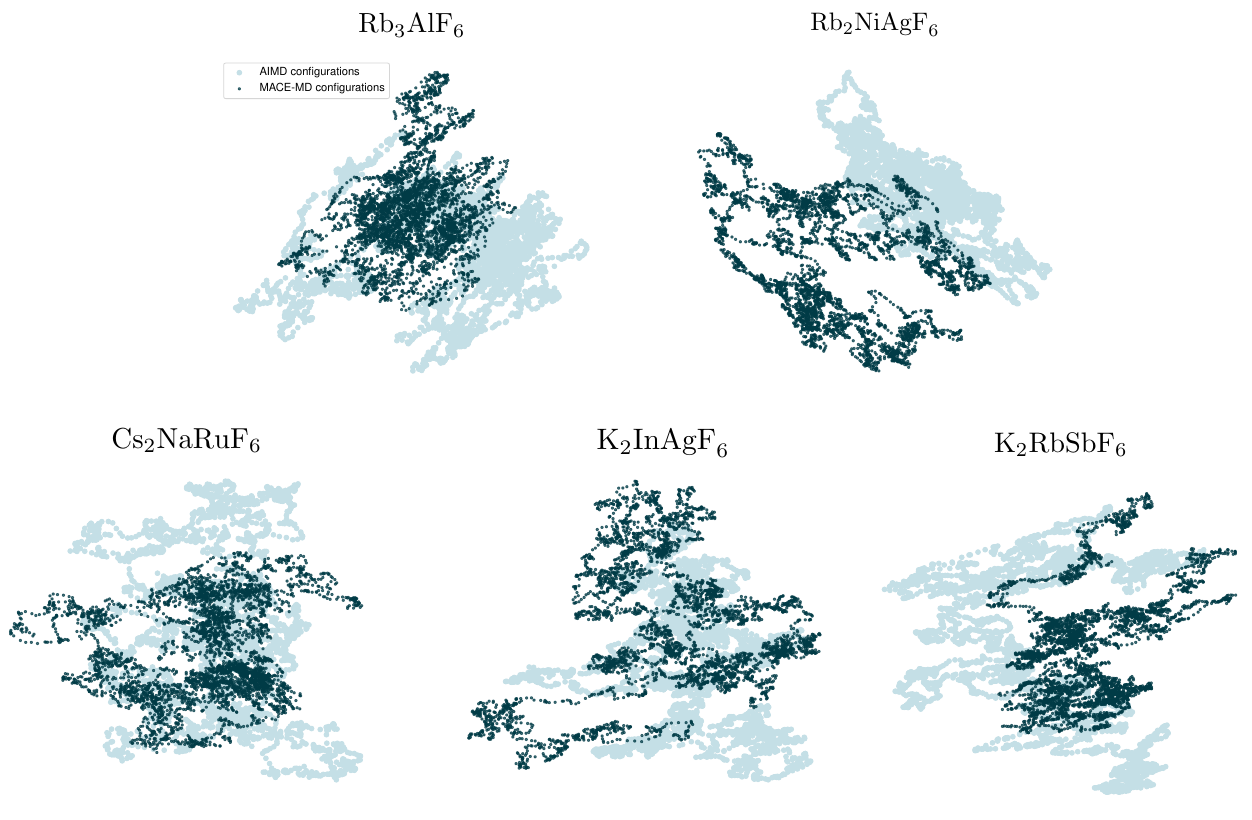}
    \caption{Same as \cref{fig:kpca_maps} which now includes a longer AIMD trajectory (4 ps in total) for each compound. }
    \label{fig:kpca_maps_long_DFT}
\end{figure}

\begin{figure}[htb]
    \centering
    \includegraphics[width=\linewidth]{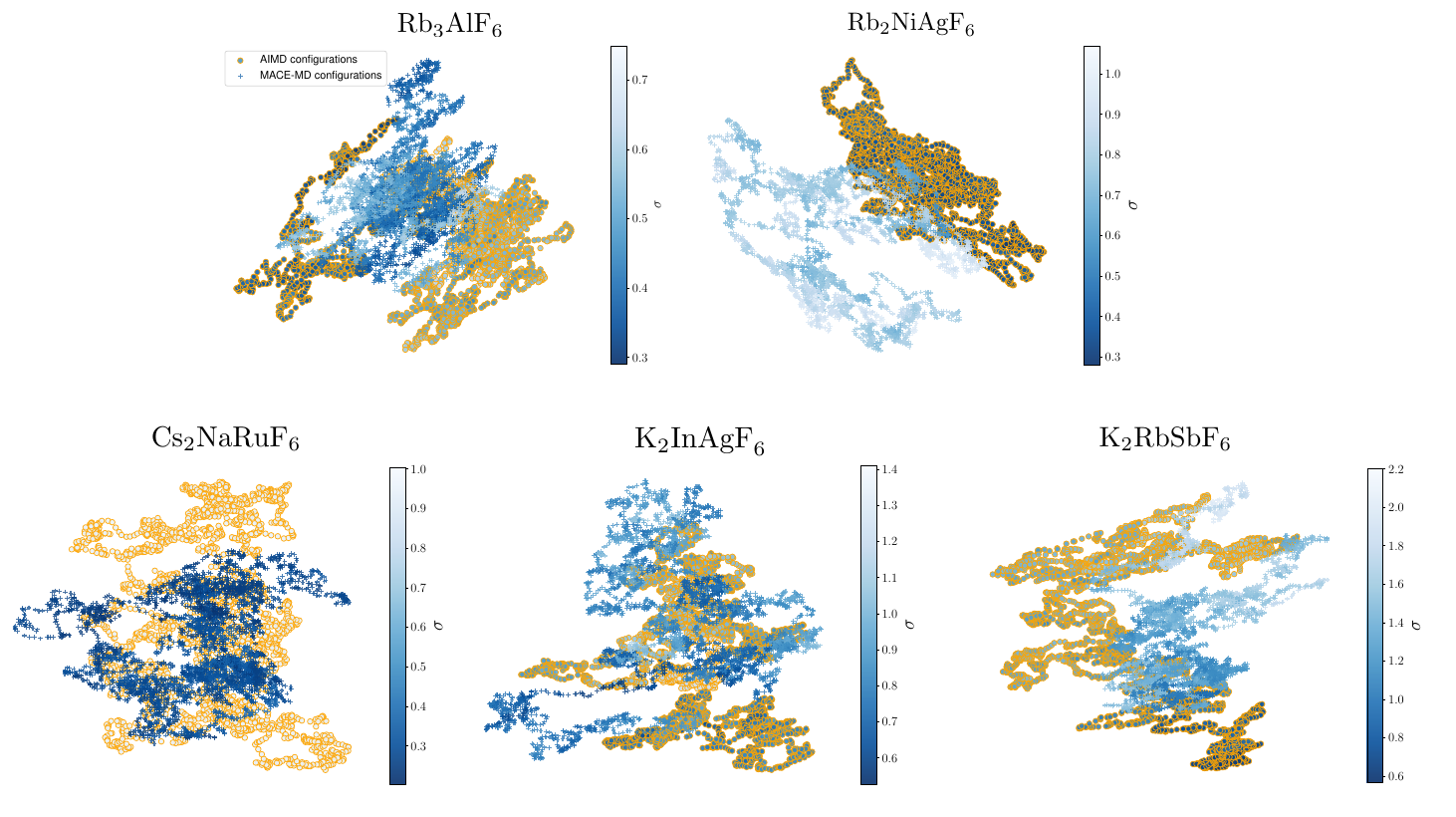}
    \caption{Same as \cref{fig:kpca_maps_sigma} which now includes a longer AIMD trajectory (4 ps in total) for each compound.}
    \label{fig:kpca_maps_long_DFT_sigma}
\end{figure}

\end{document}